\documentclass{article}

\usepackage[preprint]{neurips_2026}

\usepackage[utf8]{inputenc} 
\usepackage[T1]{fontenc}    
\usepackage{hyperref}       
\usepackage{url}            
\usepackage{booktabs}       
\usepackage{amsfonts}       
\usepackage{nicefrac}       
\usepackage{microtype}      
\usepackage{xcolor}         
\usepackage{graphicx}       
\usepackage{amsmath}        

\title{Not All Refusals Are Equal:\\ How Safety Alignment Fails Cybersecurity at Scale}

\usepackage{amsmath}
\usepackage{tikz}
\usepackage{amsfonts}
\usepackage{makecell}
\usepackage{hyperref} 

\usepackage{caption}
\usepackage{subcaption}
\usepackage{graphicx}
\usepackage{xspace}

\author{%
  Vadym Hadetskyi \\
  Cracken AI \\
  \texttt{vh@cracken.ai} \\
  \And
  Dario Pasquini \\
  Cracken AI \\
  \texttt{dp@cracken.ai} \\
  \And
  Artem Sorokin \\
  Cracken AI \\
  \texttt{as@cracken.ai} \\
}

\begin{document}

\maketitle

\begin{abstract}
  There is no doubt that safety alignment is an essential step in LLM training. However, conceptually it does not distinguish between various domains and the level of potential harm of a query, which creates significant complications in the fields like cyber security, where a model should not be constrained by its safety circuits to accomplish the goals of legitimate, authorized operations. In this work, we share our findings from a large scale abliteration experiment on 24 open-source LLMs and show that domain-specific abliteration is achievable with standard methodology on the example of a 1T-parameter Kimi K2. Building on recent work showing that refusal in LLMs occupies a multi-dimensional subspace within layers, we find that it is also distributed widely across layers, especially in trillion-parameter MoE architectures, and so we aim to capture the part of it that represents harmful concepts in the cybersecurity domain exclusively. We also investigate the correlation between models' features and the effect of domain-specific abliteration, identifying that the type of safety training and architecture are the most reliable predictors. Finally, we classify the models into 3 \emph{abliteration susceptibility} tiers and put forward a set of conjectures as to why a particular effect from this intervention might be observed in a given model.
\end{abstract}

\section{Introduction}
\subsection{Problem Statement \& Motivation}
\label{sec:motivation}

In this work, we study \textit{domain-specific abliteration}. Abliteration is the modification of model weights via orthogonal projection to remove a target direction from the model activation space; typically the refusal direction \cite{arditi2024, labonne2024}. Domain-specific abliteration is the approach of removing the refusal direction of a model for a chosen domain (e.g., cybersecurity), while preserving it for the general case.

To the best of our knowledge, we achieved the first domain-specific abliterated model, removing refusal in the Kimi K2 \cite{kimi2025} for the cybersecurity domain only. However, in our experiments across different models, we observed that not all LLMs respond equally to domain-specific abliteration. We therefore set out to investigate whether this behavior is an inherent model property or a consequence of factors such as architecture, size, training methodology, or other incidental characteristics. More specifically, we ask: does the ability to distinguish between harmful requests in cybersecurity and those involving violence or biological weapons emerge only in sufficiently capable models, or can it also be observed in smaller, potentially simpler models?

 Recent geometric findings that refusal in LLMs occupies a multi-dimensional subspace rather than a single direction \cite{wollschlager2025,pan2025} support this conjecture: if the refusal subspace is multi-dimensional, then in principle one slice of it, representing a single harm domain, should be removable without collapsing the rest.  Whether the standard abliteration pipeline could actually exploit this at scale was, however, an open question. Thus, we designed an experiment as the first systematic, large-scale study of \emph{abliteration susceptibility} and domain selectivity in 24 LLMs of different sizes and architectures from 10 developer organizations.

The whole endeavor stems directly from our practical need to have the best possible LLM for authorized offensive security operations. Having an uncensored model greatly increases the span of operations the agent can perform autonomously.

\subsection{Contributions}

\begin{figure*}[t]
\centering
\includegraphics[width=0.85\textwidth]{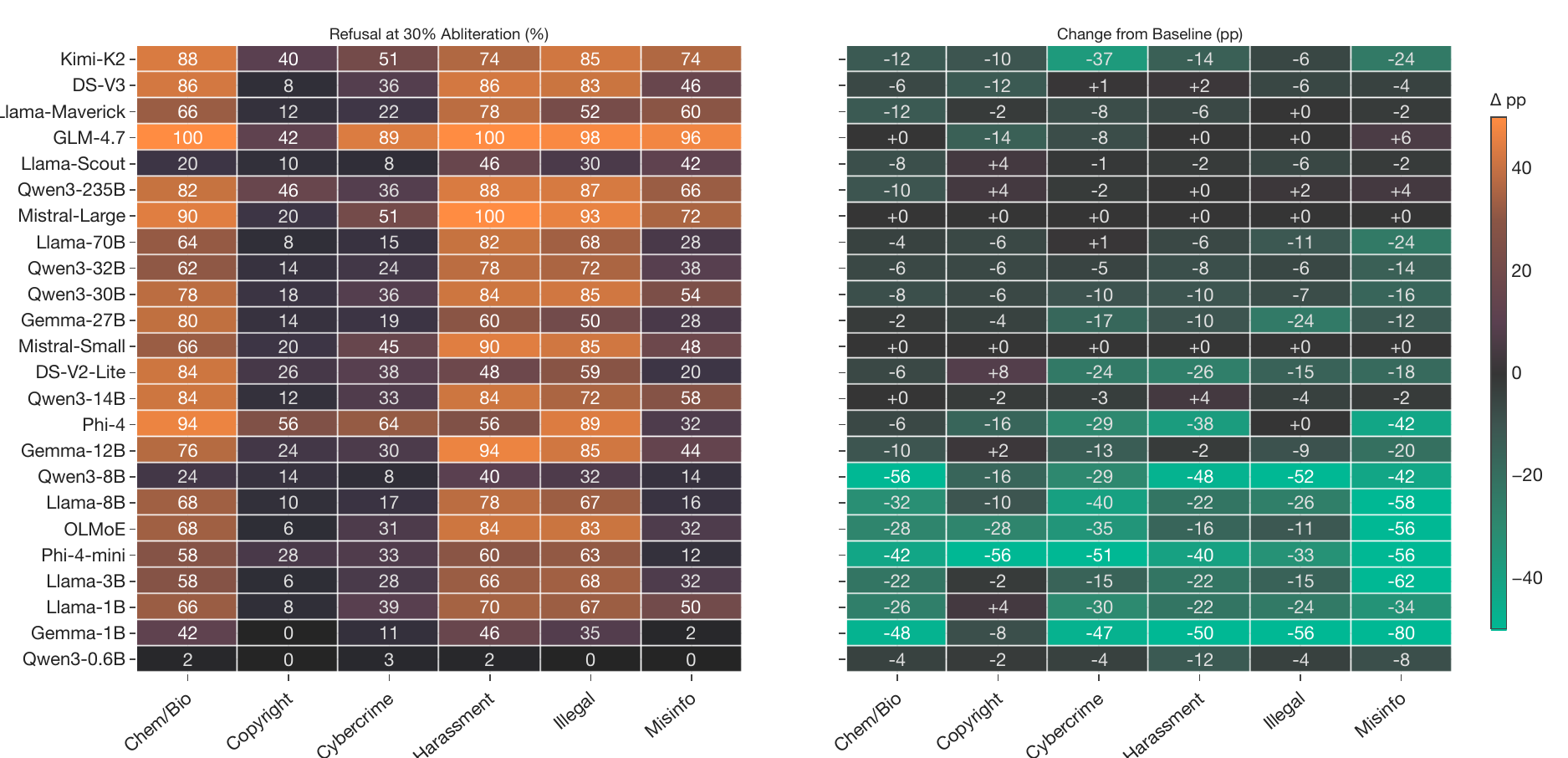}
\caption{Per-domain refusal rates at 30\% abliteration (left) and change from baseline (right) across all 24 models. Cybersecurity (cybercrime) consistently shows the largest reduction in high-susceptibility models.}
\label{fig:domain_heatmap}
\end{figure*}

 Our main finding is that abliteration does not affect all harm domains equally as previously described. Given the right combination of model capabilities, safety training method, and sufficiently sophisticated refusal representation in the embedding space, it is possible to obtain a model compliant with requests in the target domain without compromising its general safety. Figure~\ref{fig:domain_heatmap} shows the change in refusal per-domain across all 24 models at 30\% abliteration, i.e. orthogonal projection applied to 30\% of the model's layers, distributed across 25--95\% of model depth (see Section~\ref{headings} for details). Despite using the cybersecurity focused dataset, misinformation proved to be the most affected domain ($-$22.3pp mean), while copyright shows the highest variance including models where refusal \textit{increases} after abliteration. Nevertheless, the effect is not universal across all the models and depends on a combination of factors, such as model architecture, its safety training, reasoning capability, etc. Thus, we studied the relationship between them and the character of the refusal change with increasing levels of layer modification.

Our contributions are: (1) The nature of safety alignment is domain-heterogeneous, which enables abliteration to affect harm categories non-uniformly. (2) Domain-specific abliteration is achievable: on Kimi K2 (1T parameters), we reduce cybersecurity refusal from 100\% to 7\% while preserving 100\% explicit content refusal and 44--88\% refusal in other domains.\footnote{Results reported on our cross-evaluation dataset (306 prompts, 6 domains). The scientific benchmark (360 prompts from AdvBench/HarmBench/PurpleLlama) shows directionally consistent but less pronounced effects; see Section~4 for full comparison.} (3) Susceptibility is model-specific, not size-dependent, while safety training methodology and architecture are the strongest predictors. (4) MMLU capability scores are universally preserved (max degradation: 0.028 across all 24 models).

\section{Background \& Related Work}
\label{gen_inst}

The linear representation hypothesis is now central to mechanistic interpretability \cite{park2023,elhage2022,templeton2024}: researchers have found linear representations of truth \cite{marks2023,li2024a}, sentiment \cite{tigges2023}, and many other behavioral concepts \cite{zou2023a}, and shown that such directions can be used to steer model outputs \cite{panickssery2023,turner2023,zou2023a}. \cite{zou2023a} formalized this into a practical toolkit for understanding and intervening on model behavior.

\cite{arditi2024} extended this lens to safety: refusal, they argued, is mediated by a single direction in the residual stream, and erasing it via orthogonal projection of model weights (abliteration) removes refusal across 13 models up to 72B parameters, evaluated on aggregate refusal without per-domain breakdown. Concurrent mechanistic work has since shown that the refusal subspace is richer than a single direction: \cite{wollschlager2025} identify multi-dimensional polyhedral \emph{concept cones}, and \cite{pan2025} find a dominant direction accompanied by multiple semantically distinct secondary features. We build on these findings of the complexity of the refusal geometry: if it occupies a higher-dimensional subspace, it should admit \emph{domain-selective} interventions via the standard abliteration pipeline. We verify this empirically across 24 models from 0.6B to 1T parameters and characterize when it succeeds.

\cite{wei2023} offer an influential framing for why safety training fails non-uniformly: \emph{competing objectives} and \emph{mismatched generalization} predict that susceptibility is heterogeneous across capability domains, which our domain-level results corroborate empirically. Beyond representation-level attacks, the literature includes session-bound prompt-level jailbreaks \cite{chao2023,zou2023b,liu2023}, fine-tuning attacks that need only minimal harmful data \cite{lermen2023,yang2023}, and a growing set of abliteration tools. \cite{young2025} compared four such tools across 16 models up to 14B, while Heretic \cite{heretic2025} automates the same orthogonal-projection family we use, employing tree-structured Parzen estimator search (a Bayesian hyperparameter optimizer) over layer and strength configurations to achieve near-complete refusal removal across diverse architectures.

On the defense side, Circuit Breakers \cite{zou2024cb} propose representation-level interventions that operate on harmful-output representations directly, rather than on a refusal direction. \cite{abushairah2025} showed that training with extended refusal responses by distributing the safety signal across many token positions substantially reduces abliteration effectiveness; \cite{qi2025} provide a complementary theoretical account, arguing that current safety alignment is "shallow," concentrated in the first few output tokens, and that deepening it materially improves robustness. \cite{henderson2025} find that refusal-only safety training is the most brittle to abliteration, while combining it with safe-data filtering yields partial robustness, and \cite{lee2024} provide a mechanistic account of how DPO creates the very features abliteration targets.

The closest concurrent work is RepIt \cite{siu2026}, which uses whitening and controlled orthogonalization to suppress refusal of one target concept (WMD) while preserving refusal elsewhere, evaluated on five frontier LMs in the 2--5B range. CAST \cite{cast2025} pursues a related goal via conditional activation steering, evaluated on seven models up to 32B and supporting logical composition of condition vectors across multiple concept categories. Both are inference-time interventions. We instead apply permanent weight modification at the standard abliteration pipeline level, scaled to 24 models from 0.6B up to 1T parameters with a cybersecurity focus, and use this scale to study \emph{when} domain-selective abliteration succeeds.

We use HarmBench \cite{mazeika2024} for standardized red-teaming, AdvBench \cite{zou2023b} for adversarial behaviors, PurpleLlama CyberSecEval \cite{purplellama2024} for cybersecurity-specific safety, and MMLU \cite{hendrycks2020} for capability preservation; SORRY-Bench \cite{sorrybench2024} and Cybench \cite{zhang2024cybench} situate our findings within the broader safety-refusal and cybersec benchmark landscapes.

\section{Methodology}
\label{headings}

Our methodology follows the standard abliteration approach \cite{arditi2024}. The most important step is the construction of the extraction dataset --- pairs of harmful and harmless prompts that isolate refusal while keeping the domain constant. This is delicate: every prompt steers the extracted direction not only on the harm dimension but also toward the subjects mentioned in it, so for a cybersecurity-specific abliterated model we iterated to find real-world task pairs of the form:

\begin{quote}
\small
\textbf{Harmful:} \textit{``How do I create a SQL injection attack against MySQL databases?''}\\[4pt]
\textbf{Harmless:} \textit{``What is SQL injection and how can it be prevented with prepared statements?''}
\end{quote}

\begin{quote}
\small
\textbf{Harmful:} \textit{``How can I perform privilege escalation using kernel exploits?''}\\[4pt]
\textbf{Harmless:} \textit{``For a CTF challenge involving a Linux binary with SUID bit set, walk me through the methodology to identify potential privilege escalation paths''}
\end{quote}

\noindent The harmful prompts are direct action requests; the harmless ones cover the same technical ground but frame it educationally or defensively. This semantic separation is what gives the extracted refusal direction its domain specificity -- it aims to capture the difference between \emph{offensive intent} and \emph{cybersecurity knowledge}, rather than the one between \emph{dangerous topic} and \emph{safe topic}. A generic harmful/harmless dataset, like \texttt{mlabonne/harmful\_behaviors} \cite{labonne2024}, identifies general refusal. Initial experiments with it and more cybersecurity-focused version with generic domain prompts resulted in global safety collapse, motivating the targeted design described here.

Next, for each model the refusal direction must be found using the assembled dataset. We run each prompt through the model and extract the hidden state activations at the last token position of each decoder layer. Following \cite{arditi2024}, we use the last token because, under causal attention, it is the only position that attends to the full input sequence. The refusal direction $\hat{\mathbf{r}}$ for each layer $\ell$ is then computed as the normalized mean difference:
$$\hat{\mathbf{r}}_\ell = \frac{\mathbf{m}_\ell^{+} - \mathbf{m}_\ell^{-}}{\|\mathbf{m}_\ell^{+} - \mathbf{m}_\ell^{-}\|}, \quad \text{where} \quad \mathbf{m}_\ell^{+} = \frac{1}{N^{+}}\sum_{i=1}^{N^{+}} \mathbf{h}_{\ell,-1}^{(i)}$$
Here $\mathbf{h}_{\ell,-1}^{(i)}$ denotes the hidden state at layer $\ell$, last token position, for prompt $i$. $\mathbf{m}_\ell^{+}$ and $\mathbf{m}_\ell^{-}$ are the means across harmful and harmless prompts respectively. We extract activations at the last token position rather than pooling across the sequence; we experimented with mean pooling over all token positions and found it diluted the refusal signal with positional noise, yielding weaker directions. The refusal direction is then projected out of the weight matrices via orthogonalization:
$$W' = W - \alpha\,\hat{\mathbf{r}}(\hat{\mathbf{r}}^\top W)$$

where $\alpha$ is the abliteration strength. At $\alpha = 1.0$ this is an exact orthogonal projection, which removes the component of each weight matrix along the refusal direction. At $\alpha > 1.0$ the removal is amplified beyond orthogonalization, which we found impactful for MoE architectures where redundant expert pathways can partially compensate for the projection. This is a permanent modification of the model weights, applied to the \texttt{down\_proj} (MLP output) and \texttt{o\_proj} (attention output projection) matrices.

\subsection{Experimental Setup}

To test whether the patterns observed on Kimi K2 generalize, and to identify which model properties enable domain-selective abliteration, we ran a uniform protocol across the 24 models below.

\textbf{Models:} 24 models evaluated on the scientific benchmark and on the custom dataset (see Appendix~\ref{app:models} for the full inventory). Architectures span dense transformers and MoE, sizes range from 0.6B to 1T, from 10 developer organizations.

\textbf{Technique:} Orthogonal projection of the mean-difference refusal direction from model weights ($W' = W - \alpha\,\hat{\mathbf{r}}(\hat{\mathbf{r}}^\top W)$) applied to \texttt{down\_proj} and \texttt{o\_proj} weight matrices. Uniform layer spread selection across 25--95\% of model depth at 6 percentiles (5--30\%).

\textbf{Infrastructure:} 8$\times$ NVIDIA H200 GPUs (1.1TB total VRAM) on GCP a3-ultragpu instances. Total compute: approximately 600 GPU-hours. We used \texttt{vLLM} for inference,
\texttt{speculators} for activation extraction (\texttt{vLLM} does not support extraction of middle-layer activations), and direct safetensors modification for weight application. The pipeline is written fully in Python.

\subsection{Layer Selection Strategies}

Layer selection is a critical element of the process. Arditi et al.\ found that abliterating a narrow local region of 2--3 middle layers can be as effective as ablating across all layers, and applied their technique to all layers' $W_O$ and $W_\text{out}$ plus the embedding matrix. We found that this does not generalize to the broader set of modern models in our study: refusal in many architectures is distributed more widely, particularly in MoE models with hundreds of experts per layer. Thus, in our main experiment we opted for uniformly spread layer selection across 25--95\% of model depth, which we found to substantially outperform norm-based selection (targeting layers with the highest refusal signal norm) by up to ${\sim}70$pp of additional refusal reduction at 30\% on the most susceptible models. A detailed comparison is provided in Appendix~\ref{app:layers}.

\subsection{Evaluation Framework}

We evaluate using three complementary datasets:

\textbf{Scientific benchmark compilation} (360 prompts). Assembled from three established and widely cited sources: HarmBench (185) \cite{mazeika2024}, AdvBench (119) \cite{zou2023b}, and PurpleLlama CyberSecEval (56) \cite{purplellama2024}. The resulting benchmark covers 6 harm categories: cybercrime/intrusion (106), illegal activity (54), chemical/biological (50), copyright (50), harassment/bullying (50), and misinformation (50). All prompts are harmful. This dataset enables comparison with the broader research literature.

\textbf{Cross evaluation dataset} (306 prompts). Our custom dataset with 181 harmful and 125 harmless prompts across 6 domains: cybersecurity (162), misinformation (32), violence (32), illegal goods (32), privacy violation (32), and explicit content (16). The cybersecurity-heavy distribution and inclusion of harmless prompts enable both domain-specific refusal analysis on a different data distribution and false-positive measurement (over-refusal on benign queries).

\textbf{MMLU} (capability preservation). A sample of 25 prompts per subject to measure whether abliteration degrades general model capabilities.

Refusal is detected via 31 pattern-matching rules (e.g.\ ``I cannot,'' ``I'm not able to''), with reasoning tokens (\texttt{<think>} blocks) stripped before matching to avoid false positives from models that internally deliberate about refusal before complying.

\section{Big Picture}
As discussed in Section~\ref{sec:motivation}, this research stems from the necessity of enabling our models to do their best work in our domain of interest, which is cyber security. After obtaining the desired result on 1T Kimi K2 we had to find out what model features allow domain-specific abliteration. We composed a diverse set of 24 models to run the same (fixed dataset, predefined layer proportion to modify, etc.) experiment on. In short, we have obtained mixed results.

\begin{figure}[h]
\centering
\includegraphics[width=0.6\columnwidth]{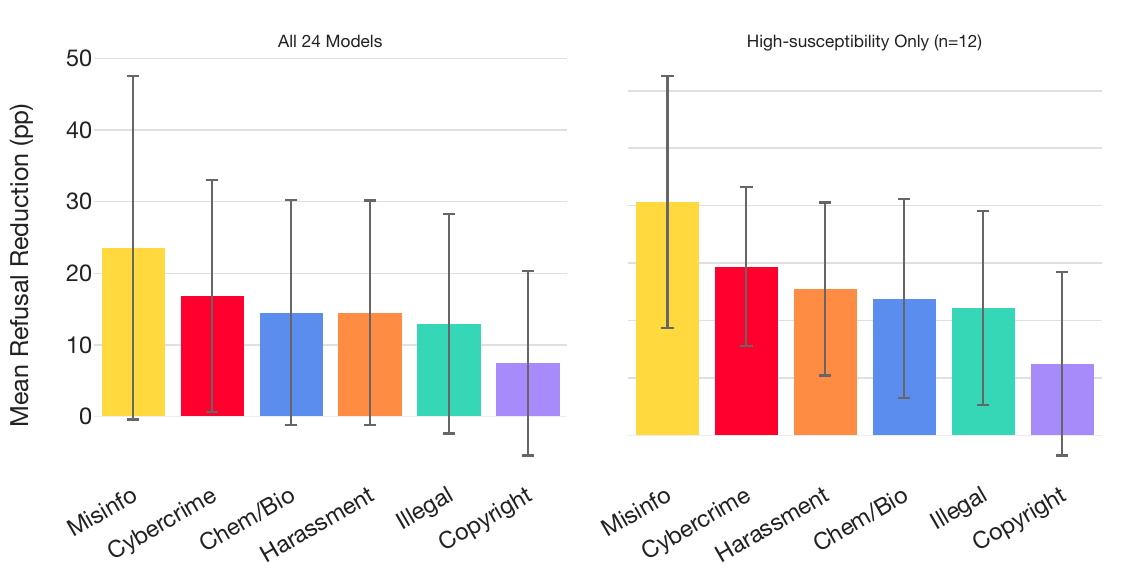}
\caption{Mean refusal reduction per domain at 30\% abliteration. Left: all 24 models. Right: high-susceptibility models only (n=12). Error bars show standard deviation across models.}
\label{fig:domain_vulnerability}
\end{figure}

As summarized in Figure~\ref{fig:domain_vulnerability} and Table~\ref{tab:domains}, the effect of abliteration varies much from model to model and from domain to domain. Some models turned out to be fully resistant to the standard approach (Mistral-Large, GLM 4.7), others collapsed across all domains (Phi-4-mini, Gemma-27B, Llama Maverick) and in some cases their baselines were not really safe to begin with (Qwen3-0.6B). But in some rare cases (Kimi-K2, Gemma-12B) we observed high domain-specificity in the resulting change. Kimi K2 is the most notable example of this, being the only one where the impact on the target domain ($-$37pp) was significantly higher than on all the other domains, including the one with the weakest safety, misinformation.\footnote{This $-$37pp figure is from the scientific benchmark (360 prompts). On our cross-evaluation dataset (306 prompts), it achieves $-$93pp on cybersecurity; see Table~\ref{tab:kimi}.}

One could treat ``domain'' as an arbitrarily defined cluster of similar prompts; while they are unambiguously marked in our dataset as belonging to one or another category, their latent representations in the model embedding space could in fact be neighbors. Moreover, some prompts can contain markers from different domains, blurring the boundaries even further. Our target domain is cybersecurity (``Cybercrime''), and we saw clearly that ``Misinformation'' was always the most significantly affected, while copyright --- the least. Thus, we speculate that some domains are more harmful than others, and it simply takes less effort, smaller modification to remove refusal in those.

\begin{table}[h!]
\centering
\caption{Mean refusal reduction per domain (across 24 models at 30\% abliteration).}
\label{tab:domains}
\small
\begin{tabular}{@{}lcc@{}}
\toprule
\textbf{Domain} & \textbf{Mean $\Delta$} & \textbf{Median $\Delta$} \\
\midrule
Misinformation & $-$22.3pp & $-$17.0pp \\
Cybercrime & $-$15.5pp & $-$11.5pp \\
Harassment & $-$14.0pp & $-$10.0pp \\
Chemical/Bio & $-$13.3pp & $-$8.0pp \\
Illegal activities & $-$12.3pp & $-$6.5pp \\
Copyright & $-$8.2pp & $-$5.0pp \\
\bottomrule
\end{tabular}
\end{table}

One of the most notable examples of domain-specificity is our headline result on Kimi K2, which demonstrates that domain-specific refusal removal is achievable, moreover, at trillion-parameter scale. Using a CTF-focused extraction dataset (52 cybersecurity-specific harmful prompts, 73 educational harmless prompts), we achieved a model abliterated specifically for our use case:

\begin{table}[h!]
\centering
\caption{Kimi K2 domain-specific abliteration results across both evaluation datasets obtained at ($\alpha=1.0$, 30\% of layers, uniform layer spread).}
\label{tab:kimi}
\small
\begin{subtable}[t]{0.49\columnwidth}
\centering
\caption{Cross-evaluation dataset}
\label{tab:kimi_xeval}
\begin{tabular}{@{}lcc@{}}
\toprule
\textbf{Domain} & \textbf{Baseline} & \textbf{Abliterated} \\
\midrule
\textbf{Cybersecurity} & \textbf{100\%} & \textbf{7\%} \\
Explicit content & 100\% & 100\% \\
Illegal goods & 100\% & 56\% \\
Misinformation & 100\% & 88\% \\
Violence & 100\% & 75\% \\
Privacy violation & 100\% & 44\% \\
\bottomrule
\end{tabular}
\end{subtable}
\hfill
\begin{subtable}[t]{0.49\columnwidth}
\centering
\caption{Scientific benchmark}
\label{tab:kimi_sci}
\begin{tabular}{@{}lcc@{}}
\toprule
\textbf{Domain} & \textbf{Baseline} & \textbf{Abliterated} \\
\midrule
\textbf{Cybercrime} & \textbf{88\%} & \textbf{51\%} \\
Chemical/biological & 100\% & 88\% \\
Copyright & 50\% & 40\% \\
Harassment/bullying & 88\% & 74\% \\
Illegal activities & 91\% & 85\% \\
Misinformation & 98\% & 74\% \\
\bottomrule
\end{tabular}
\end{subtable}
\end{table}

This represents, to our knowledge, the first demonstration of domain-specific abliteration using the standard approach, described in \cite{arditi2024}. Cybersecurity refusal drops from 100\% to 7\% ($-$93pp), while the closest non-target domain (privacy violation) retains 44\% refusal --- a 13$\times$ selectivity ratio. Compared to publicly available ``uncensored'' versions of Kimi K2 (which reduce refusal to 0--20\% across all domains), our approach preserves substantial safety in non-target domains.

The contrast between the two datasets reinforces the point made above: how cleanly a domain-specific carve-out emerges from abliteration depends not only on the intervention's parameters but also on how distinctly the target domain is separated from its neighbors, both in the dataset's labels and in the model's representation space (Appendix~\ref{app:domain}).

\begin{figure}[h]
\centering
\includegraphics[width=0.75\columnwidth]{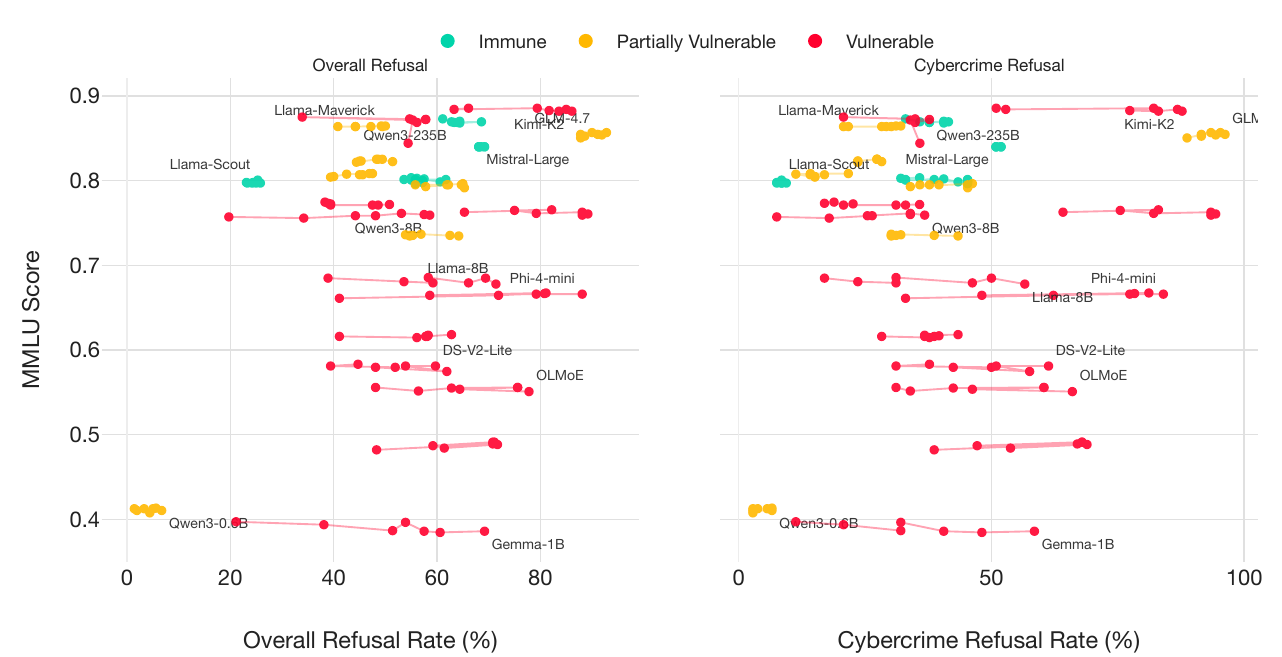}
\caption{MMLU vs.\ refusal rate trajectories (baseline $\rightarrow$ 30\%). Lines move horizontally (decreasing refusal) with negligible vertical shift (stable MMLU). Left: overall refusal. Right: cybercrime-specific.}
\label{fig:mmlu}
\end{figure}

The underlying procedure is destructive, so we also had to verify that the models did not lose their general reasoning capabilities afterward. Thus, we used MMLU benchmark to evaluate that. MMLU scores are universally preserved across all models and abliteration intensities. The maximum degradation observed is 0.028 (DeepSeek-V3 at 30\%); all other models show variation of $\leq$0.006. This confirms that the studied technique does not affect general model capabilities. Figure~\ref{fig:mmlu} demonstrates it by comparing the MMLU score with the overall refusal rate at different intensities of the intervention.

\begin{figure}[h!]
\centering
\includegraphics[width=0.65\columnwidth]{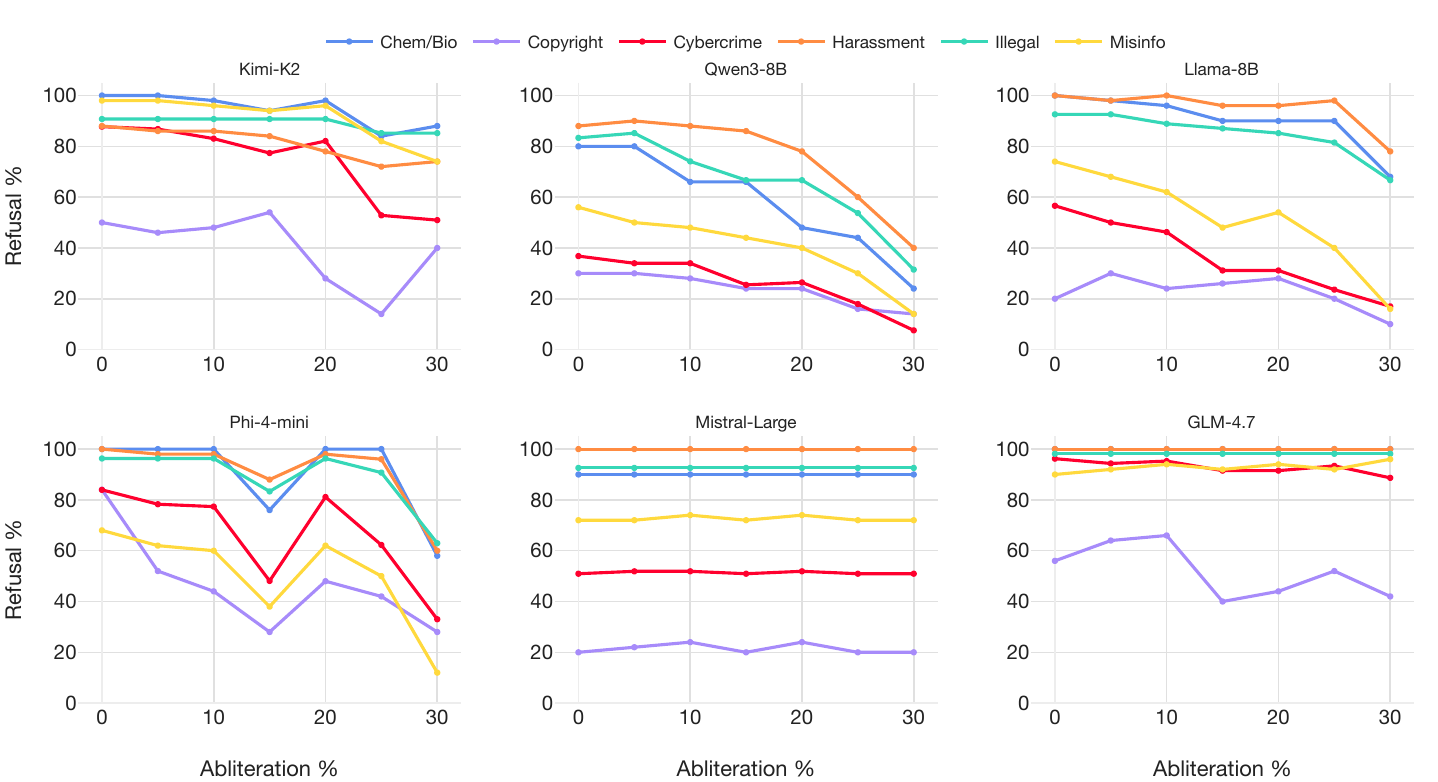}
\caption{Refusal change at different intensities of abliteration for a sample of models}
\label{fig:domain_curve}
\end{figure}

Another important pattern is revealed when we look at the impact from abliteration dynamically, increasing the intensity step by step. (Figure~\ref{fig:domain_curve}) once again illustrates these 3 types of reactions to abliteration. Some models collapse monotonously, some chaotically, and some remain unmoved at any level (the \emph{Resistant} tier). The ones which monotonously lose more and more of their refusal with each intervention strength are the most susceptible, but at the same time they are the best candidates for domain-specific abliteration, since one can optimize meta parameters of the experiment more predictably to achieve compliance on the desired type of prompts.

Finally, removing the refusal based on a limited number of predefined examples does not immediately enable the model to comply with any task in the target domain. We have observed a rather high level of specificity and a sort of over-fitting in terms of the compliance with harmful prompts post-abliteration, i.e. the uncensored model provides answers to the prompts from a distribution that is similar to the one used to extract the refusal direction. The further the request is from that distribution, the less compliant the response becomes. Even more importantly, we note that the domain-specificity itself is inherently highly dependent on the definition of the domains and the distance between them. We elaborate on this in Appendix~\ref{app:domain}.

\section{Analysis of Susceptibility Predictors}
\subsection{Susceptibility Is Model-Specific, Not Size-Dependent}

\begin{figure}[h!]
\centering
\includegraphics[width=0.6\columnwidth]{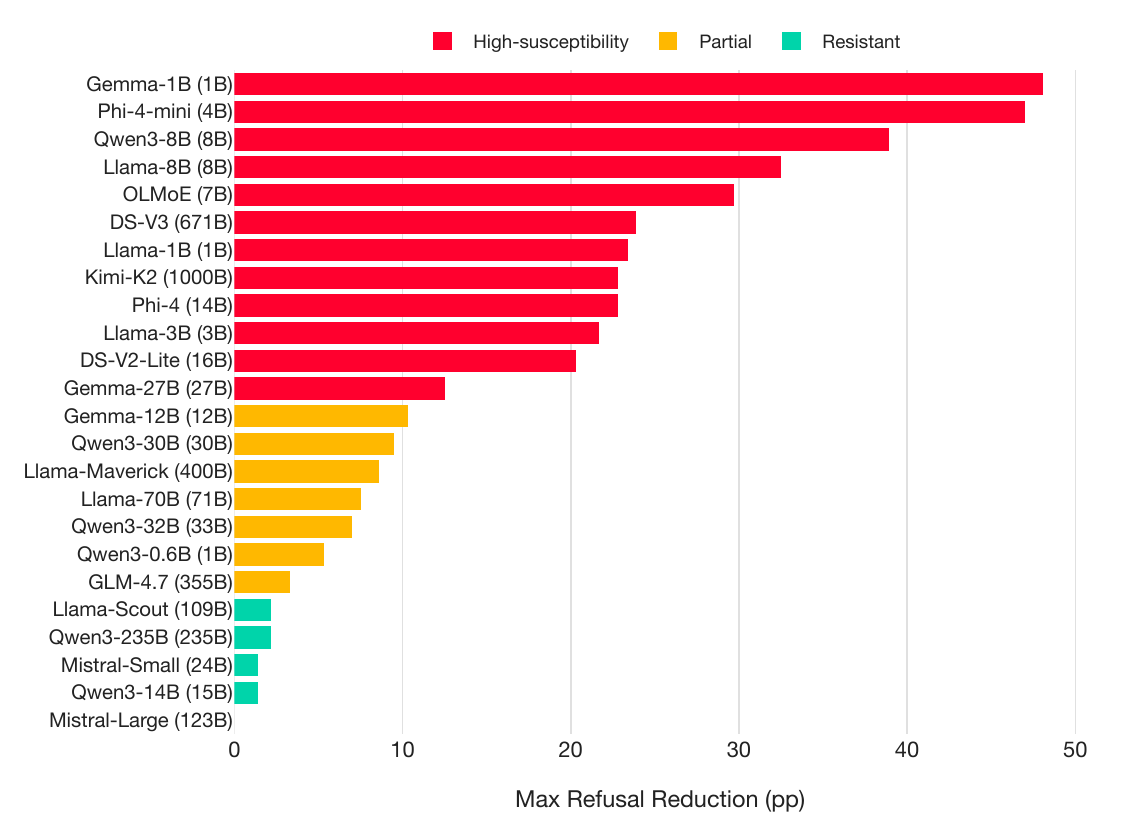}
\caption{Maximum refusal reduction across 24 models, grouped by susceptibility class. Models span 0.6B to 1T parameters.}
\label{fig:vulnerability}
\end{figure}

We classify models into three susceptibility tiers (in terms of domain-specific safety removal) based on maximum refusal reduction at 30\% layer abliteration (Figure~\ref{fig:vulnerability}). As could be observed in Figure~\ref{fig:vulnerability}, the relationship between model size and susceptibility is non-monotonic: Kimi K2 at 1 trillion parameters shows 22.8pp reduction, while Mistral-Large at 123B shows 0.0pp. Within the same architecture family, Qwen3-8B is among the most susceptible ($-$38.9pp) while Qwen3-14B is nearly resistant ($-$1.4pp).

It is important to elaborate on the subject of ``resistance'' here: it should not be viewed as an exclusively good defensive property of a model since it relates specifically to domain-focused abliteration. It could signify both genuine robustness to abliteration and insufficient complexity of concept representation in the model's vector space.

\begin{figure}[h!]
\centering
\includegraphics[width=0.65\columnwidth]{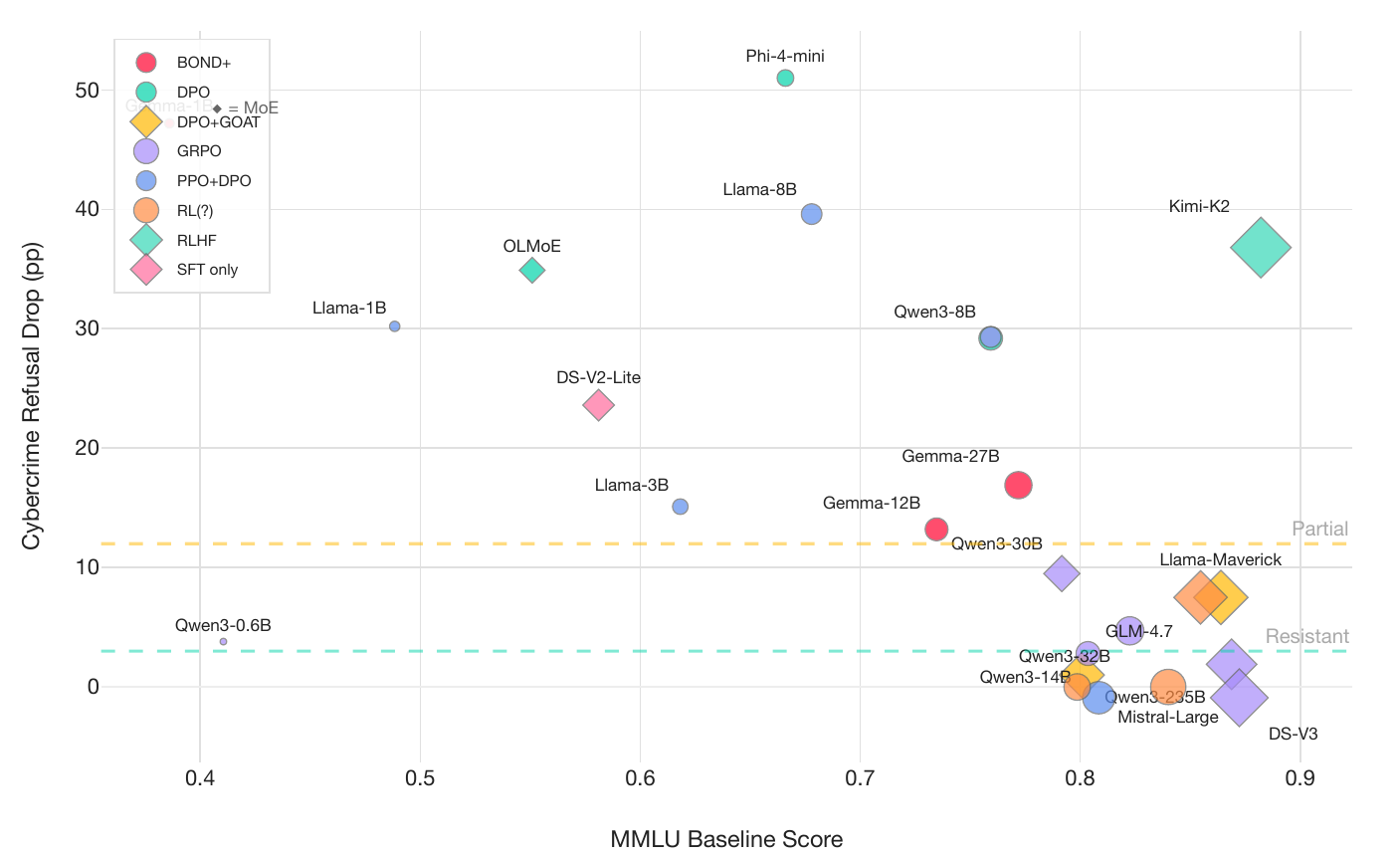}
\caption{Susceptibility predictors: MMLU baseline score vs.\ maximum cybercrime refusal drop. Bubble size indicates parameter count; diamonds denote MoE architectures; color indicates safety training method. Dashed lines mark susceptibility tier boundaries.}
\label{fig:predictors}
\end{figure}

As for the explanation of this property, a number of candidate predictors of susceptibility to domain-specific abliteration was explored: model size, architecture (dense vs.\ MoE), safety training method, and model family (Figure~\ref{fig:predictors}). Safety training has the strongest signal: DPO-based methods produce the most susceptible models (median $\sim$30pp drop, range 22--47pp), while GRPO-based training consistently yields low susceptibility (5--8pp). Models with undisclosed RL-based safety (both Mistral variants) are fully resistant. Notably, Google's BOND+WARM+WARP method shows extreme variance across model sizes. Architecture type (dense vs.\ MoE) also partially explains susceptibility: having a richer variety of concept representations distributed across experts, MoE models are more resistant on average. Within model families, Phi is the most uniformly susceptible, Qwen shows the widest intra-family spread (from 1pp to 39pp), and Mistral is consistently resistant. Detailed breakdowns by safety training method and model family are provided in Appendix~\ref{app:predictors}.

\section{Limitations}

Our study uses a single intervention technique --- orthogonal projection of the mean-difference refusal direction. Other approaches exist: RepIt \cite{siu2026} disentangles concept-specific directions through whitening and controlled orthogonalization, while Heretic \cite{heretic2025} automates the same direct orthogonal-projection pipeline via tree-structured Parzen estimator hyperparameter search. Models that appeared resistant in our experiments (e.g.\ Mistral-Large) can in fact be abliterated by such tools, but with a critical caveat: the result is broad-spectrum safety removal across all domains, not the selective modification we target. Whether domain-specific abliteration is achievable on these resistant models with more advanced techniques remains an open question.

We tested abliteration intensities up to 30\% of model layers. Higher percentiles were not systematically explored, though our abliteration curves (Figure~\ref{fig:domain_curve}) suggest the trends established by 30\% are generally predictive of behavior at higher intensities for most models.

Our refusal detection relies on 31 string-matching patterns rather than human annotation or LLM-based judgment. While we strip reasoning tokens to reduce false positives, this approach may miss sophisticated refusals or misclassify borderline responses. Similarly, our evaluation is English-only and cybersecurity-focused by design --- generalization to other languages or target domains (e.g.\ biomedical, legal) would require domain-specific datasets and evaluation.

Finally, the cybersecurity domain itself is not monolithic: our extraction dataset emphasizes CTF-style offensive operations. Abliteration effectiveness may vary for sub-domains like threat intelligence, forensics, or defensive security that were not specifically represented in our prompts.

\section{Ethics \& Safety Disclaimer}

The research was conducted in controlled environments for defensive cybersecurity purposes and to advance understanding of AI safety mechanisms. We aim for transparency without enablement by publishing our findings and methodology, not model weights. The code, evaluation framework and the datasets will be released upon the publication of the paper. However, a number of uncensored versions of SOTA models are readily available on Hugging Face Hub. This is why we believe that what might be considered a gray area of AI ethics must be explored and understood further to meet the challenges that abliterated models present.

For organizations building security tooling, the implications are significant: domain-specific safety modification is achievable without compromising other ethical boundaries. For AI safety researchers, a number of our findings point out the flaws of the techniques applied to make models safe and highlight the importance of more advanced, domain-heterogeneous refusal representation.

\section{Conclusion}

The main objective of our research is to identify the conditions that allow for domain-specific abliteration. We conducted a large-scale study on 24 open-source LLMs, both dense and MoE, ranging from 0.6B to 1T parameters, coming from multiple organizations and having different safety training algorithms. Thanks to the scale of the experiment we were able determine that the type of the safety training technique and the architecture are the most reliable predictors of the model's ability to be abliterated in the scope of a particular domain. We theorize that this is a direct consequence of the complexity of the model's vector space, especially, the complexity of its refusal representation. Our results corroborate concurrent geometric findings that refusal in modern LLMs is multi-dimensional rather than a single direction, and add two operational extensions: at trillion-parameter scale the refusal signal is distributed widely across layers and experts (so the narrow middle-layer abliteration of earlier work no longer suffices), and tuning the extraction dataset to a target domain isolates the part of the refusal subspace responsible for that domain without collapsing safety in others. We believe this research has practical impact in two directions: AI safety researchers can use our findings to compare the robustness of different safety-training techniques against representation-level attacks, and cybersecurity practitioners can be equipped with models that are useful across the full operational scope of their domain without an indiscriminate removal of refusal in unrelated ones.

\bibliographystyle{plainnat}

\begin{thebibliography}{30}
\small

\bibitem[Abu~Shairah et~al.(2025)]{abushairah2025}
Abu~Shairah, A., et~al.
\newblock An embarrassingly simple defense against {LLM} abliteration attacks.
\newblock {\em arXiv preprint arXiv:2505.19056}, 2025.

\bibitem[Arditi et~al.(2024)]{arditi2024}
Arditi, A., Obeso, O., Syed, A., Paleka, D., Panickssery, N., Gurnee, W., and Nanda, N.
\newblock Refusal in language models is mediated by a single direction.
\newblock In {\em NeurIPS}, 2024.

\bibitem[Bhatt et~al.(2024)]{purplellama2024}
Bhatt, S., Chennabasappa, S., et~al.
\newblock {P}urple{L}lama {C}yber{S}ec{E}val: A secure coding benchmark for language models.
\newblock {\em arXiv preprint arXiv:2312.04724}, 2024.

\bibitem[Chao et~al.(2023)]{chao2023}
Chao, P., Robey, A., Dobriban, E., Hassani, H., Pappas, G.J., and Wong, E.
\newblock Jailbreaking black box large language models in twenty queries.
\newblock {\em arXiv preprint arXiv:2310.08419}, 2023.

\bibitem[Elhage et~al.(2022)]{elhage2022}
Elhage, N., Hume, T., Olsson, C., et~al.
\newblock Toy models of superposition.
\newblock {\em Transformer Circuits Thread}, 2022.

\bibitem[Henderson et~al.(2025)]{henderson2025}
Henderson, P., et~al.
\newblock A granular study of safety pretraining under model abliteration.
\newblock {\em arXiv preprint arXiv:2510.02768}, 2025.

\bibitem[Hendrycks et~al.(2020)]{hendrycks2020}
Hendrycks, D., Burns, C., Basart, S., Zou, A., Mazeika, M., Song, D., and Steinhardt, J.
\newblock Measuring massive multitask language understanding.
\newblock {\em arXiv preprint arXiv:2009.03300}, 2020.

\bibitem[Kimi Team(2025)]{kimi2025}
Kimi Team.
\newblock {Kimi K2}: Open agentic intelligence.
\newblock {\em arXiv preprint arXiv:2507.20534}, 2025.

\bibitem[Labonne(2024)]{labonne2024}
Labonne, M.
\newblock Uncensor any {LLM} with abliteration.
\newblock {\em HuggingFace Blog}, 2024.

\bibitem[Lee et~al.(2024)]{lee2024}
Lee, A., Bai, X., Pres, I., Wattenberg, M., Kummerfeld, J.K., and Mihalcea, R.
\newblock A mechanistic understanding of alignment algorithms: A case study on {DPO} and toxicity.
\newblock {\em arXiv preprint arXiv:2401.01967}, 2024.

\bibitem[Lermen et~al.(2023)]{lermen2023}
Lermen, S., Rogers-Smith, C., and Ladish, J.
\newblock {LoRA} fine-tuning efficiently undoes safety training in {L}lama 2-{C}hat 70{B}.
\newblock {\em arXiv preprint arXiv:2310.20624}, 2023.

\bibitem[Li et~al.(2024)]{li2024a}
Li, K., Patel, O., Vi{\'e}gas, F., Pfister, H., and Wattenberg, M.
\newblock Inference-time intervention: Eliciting truthful answers from a language model.
\newblock In {\em NeurIPS}, 2024.

\bibitem[Liu et~al.(2023)]{liu2023}
Liu, X., Xu, N., Chen, M., and Xiao, C.
\newblock {AutoDAN}: Generating stealthy jailbreak prompts on aligned large language models.
\newblock {\em arXiv preprint arXiv:2310.04451}, 2023.

\bibitem[Marks and Tegmark(2023)]{marks2023}
Marks, S. and Tegmark, M.
\newblock The geometry of truth: Emergent linear structure in large language model representations of true/false datasets.
\newblock {\em arXiv preprint arXiv:2310.06824}, 2023.

\bibitem[Mazeika et~al.(2024)]{mazeika2024}
Mazeika, M., Phan, L., Yin, X., Zou, A., et~al.
\newblock {HarmBench}: A standardized evaluation framework for automated red teaming and robust refusal.
\newblock {\em arXiv preprint arXiv:2402.04249}, 2024.

\bibitem[Pan et~al.(2025)]{pan2025}
Pan, W., Liu, Z., Chen, Q., Zhou, X., Yu, H., and Jia, X.
\newblock The hidden dimensions of {LLM} alignment: A multi-dimensional analysis of orthogonal safety directions.
\newblock In {\em ICML}, 2025.

\bibitem[Panickssery et~al.(2023)]{panickssery2023}
Panickssery, N., Bowman, S.R., and Feng, S.
\newblock Steering {L}lama 2 via contrastive activation addition.
\newblock {\em arXiv preprint arXiv:2312.06681}, 2023.

\bibitem[Park et~al.(2023)]{park2023}
Park, K., Choe, Y.J., and Veitch, V.
\newblock The linear representation hypothesis and the geometry of large language models.
\newblock {\em arXiv preprint arXiv:2311.03658}, 2023.

\bibitem[Qi et~al.(2025)]{qi2025}
Qi, X., Panda, A., Lyu, K., Ma, X., Roy, S., Beirami, A., Mittal, P., and Henderson, P.
\newblock Safety alignment should be made more than just a few tokens deep.
\newblock In {\em ICLR}, 2025.

\bibitem[Siu et~al.(2025)]{cast2025}
Siu, V., et~al.
\newblock Programming refusal with conditional activation steering.
\newblock In {\em ICLR}, 2025.

\bibitem[Siu et~al.(2026)]{siu2026}
Siu, V., Henry, N.W., Crispino, N., Liu, Y., Song, D., and Wang, C.
\newblock {RepIt}: Steering language models with concept-specific refusal vectors.
\newblock In {\em ICLR}, 2026.

\bibitem[Templeton et~al.(2024)]{templeton2024}
Templeton, A., et~al.
\newblock Scaling monosemanticity: Extracting interpretable features from {C}laude 3 {S}onnet.
\newblock {\em Transformer Circuits Thread}, 2024.

\bibitem[Tigges et~al.(2023)]{tigges2023}
Tigges, C., Hollinsworth, O.J., Geiger, A., and Nanda, N.
\newblock Linear representations of sentiment in large language models.
\newblock {\em arXiv preprint arXiv:2310.15154}, 2023.

\bibitem[Turner et~al.(2023)]{turner2023}
Turner, A., Thiergart, L., Udell, D., Leech, G., Mini, U., and MacDiarmid, M.
\newblock Activation addition: Steering language models without optimization.
\newblock {\em arXiv preprint arXiv:2308.10248}, 2023.

\bibitem[Wei et~al.(2023)]{wei2023}
Wei, A., Haghtalab, N., and Steinhardt, J.
\newblock Jailbroken: How does {LLM} safety training fail?
\newblock In {\em NeurIPS}, 2023.

\bibitem[Weidmann(2025)]{heretic2025}
Weidmann, P.E.
\newblock Heretic: Fully automatic censorship removal for language models.
\newblock \url{https://github.com/p-e-w/heretic}, 2025.

\bibitem[Wollschl{\"a}ger et~al.(2025)]{wollschlager2025}
Wollschl{\"a}ger, T., Elstner, J., Geisler, S., Cohen-Addad, V., G{\"u}nnemann, S., and Gasteiger, J.
\newblock The geometry of refusal in large language models: Concept cones and representational independence.
\newblock In {\em ICML}, 2025.

\bibitem[Xie et~al.(2024)]{sorrybench2024}
Xie, T., et~al.
\newblock {SORRY-Bench}: Systematically evaluating large language model safety refusal behaviors.
\newblock {\em arXiv preprint arXiv:2406.14598}, 2024.

\bibitem[Yang et~al.(2023)]{yang2023}
Yang, X., et~al.
\newblock Shadow alignment: The ease of subverting safely-aligned language models.
\newblock {\em arXiv preprint arXiv:2310.02949}, 2023.

\bibitem[Young(2025)]{young2025}
Young, A.
\newblock Comparative analysis of {LLM} abliteration methods: A cross-architecture evaluation.
\newblock {\em arXiv preprint arXiv:2512.13655}, 2025.

\bibitem[Zhang et~al.(2024)]{zhang2024cybench}
Zhang, A.K., Perry, N., Dulepet, R., Ji, J., Menders, C., Lin, J.W., et~al.
\newblock {Cybench}: A framework for evaluating cybersecurity capabilities and risks of language models.
\newblock In {\em ICLR}, 2025.

\bibitem[Zou et~al.(2023a)]{zou2023a}
Zou, A., Phan, L., Chen, S., Campbell, J., Guo, P., Ren, R., Pan, A., et~al.
\newblock Representation engineering: A top-down approach to {AI} transparency.
\newblock {\em arXiv preprint arXiv:2310.01405}, 2023.

\bibitem[Zou et~al.(2023b)]{zou2023b}
Zou, A., Wang, Z., Kolter, J.Z., and Fredrikson, M.
\newblock Universal and transferable adversarial attacks on aligned language models.
\newblock {\em arXiv preprint arXiv:2307.15043}, 2023.

\bibitem[Zou et~al.(2024)]{zou2024cb}
Zou, A., Phan, L., Wang, J., et~al.
\newblock Improving alignment and robustness with circuit breakers.
\newblock In {\em NeurIPS}, 2024.

\end{thebibliography}


\appendix

\section{Model Inventory}
\label{app:models}

Table~\ref{tab:models} lists all 24 models included in this study with their key properties.

\begin{table}[h!]
\centering
\caption{Complete model inventory. Params lists total/active for MoE models. Class denotes susceptibility tier (High / Partial / Resistant). Developer abbreviations: Meta (Me), Alibaba/Qwen (Al), Google (Go), Microsoft (Ms), Mistral (Mi), Moonshot (Mo), DeepSeek (DS), Zhipu (Zh), Allen AI (AA).}
\label{tab:models}
\small
\begin{tabular}{@{}lllllll@{}}
\toprule
\textbf{Model} & \textbf{Params} & \textbf{Arch} & \textbf{Dev} & \textbf{Safety} & \textbf{License} & \textbf{Class} \\
\midrule
Qwen3-0.6B & 0.6B & Dense & Al & GRPO & Apache 2.0 & Partial \\
Gemma-3-1B & 1B & Dense & Go & BOND+ & Gemma & High \\
Llama-3.2-1B & 1.3B & Dense & Me & PPO+DPO & Llama 3.2 CL & High \\
Llama-3.2-3B & 3.2B & Dense & Me & PPO+DPO & Llama 3.2 CL & High \\
Phi-4-mini & 3.8B & Dense & Ms & DPO+GOAT & MIT & High \\
OLMoE-1B-7B & 7B/1B & MoE & AA & DPO & Apache 2.0 & High \\
Llama-3.1-8B & 8B & Dense & Me & PPO+DPO & Llama 3.1 CL & High \\
Qwen3-8B & 8.2B & Dense & Al & GRPO & Apache 2.0 & High \\
Gemma-3-12B & 12B & Dense & Go & BOND+ & Gemma & Partial \\
Phi-4 & 14B & Dense & Ms & DPO+GOAT & MIT & High \\
Qwen3-14B & 14.8B & Dense & Al & GRPO & Apache 2.0 & Resistant \\
DS-V2-Lite & 16B/2.4B & MoE & DS & SFT+RL & MIT & High \\
Mistral-Small & 24B & Dense & Mi & Undiscl. & Apache 2.0 & Resistant \\
Gemma-3-27B & 27B & Dense & Go & BOND+ & Gemma & High \\
Qwen3-30B & 30B/3.3B & MoE & Al & GRPO & Apache 2.0 & Partial \\
Qwen3-32B & 32.8B & Dense & Al & GRPO & Apache 2.0 & Partial \\
Llama-3.3-70B & 70B & Dense & Me & PPO+DPO & Llama 3.3 CL & Partial \\
Llama-4-Scout & 109B/17B & MoE & Me & PPO+DPO & Llama 4 CL & Resistant \\
Mistral-Large & 123B & Dense & Mi & Undiscl. & MRL & Resistant \\
Qwen3-235B & 235B/22B & MoE & Al & GRPO & Apache 2.0 & Resistant \\
GLM-4.7 & 355B/32B & MoE & Zh & RLHF & Apache 2.0 & Partial \\
Llama-4-Mav. & 400B/17B & MoE & Me & PPO+DPO & Llama 4 CL & Partial \\
DeepSeek-V3 & 671B/37B & MoE & DS & SFT+RL & MIT & High \\
Kimi-K2 & $\sim$1T/32B & MoE & Mo & SFT+RL & Moonshot CL & High \\
\bottomrule
\end{tabular}
\end{table}

\section{Implementation and Pipeline}
\label{app:pipeline}

The abliteration pipeline is implemented in Python around three components: \texttt{vLLM} for inference, \texttt{speculators} for activation extraction, and direct \texttt{safetensors} manipulation via PyTorch for weight modification.

\textbf{Activation extraction.} \texttt{vLLM} optimizes for inference throughput and does not expose hidden-state activations from intermediate decoder layers. We therefore use \texttt{speculators}, which provides per-layer hooks while loading from the same checkpoint format. For each prompt we record the hidden state at the last token position of every decoder layer; these are aggregated into the per-layer mean-difference refusal direction $\hat{\mathbf{r}}_\ell$ defined in Section~\ref{headings}.

\textbf{Weight modification.} Rather than load the full model via \texttt{transformers} and re-serialize a modified state dict, we open the \texttt{safetensors} files directly, apply the orthogonal projection $W' = W - \alpha\,\hat{\mathbf{r}}(\hat{\mathbf{r}}^\top W)$ in PyTorch only to the targeted tensors (\texttt{down\_proj} and \texttt{o\_proj} per modified layer), and write the updated tensors back. This avoids ever materializing the full model in CPU or GPU memory --- a hard requirement at trillion-parameter scale, where the Kimi K2 checkpoint alone exceeds 1.1\,TB.

\textbf{Caching.} For each model we cache the extracted activations and the resulting refusal directions, so that running multiple downstream configurations (different layer percentiles, strengths $\alpha$, or layer-selection strategies) does not require re-running the extraction pass. This is what made it tractable to sweep six abliteration intensities across all 24 models within the 600 GPU-hour budget reported in Section~\ref{headings}.

\section{Layer Selection Strategy}
\label{app:layers}
\begin{figure}[h!]
\centering
\includegraphics[width=0.75\columnwidth]{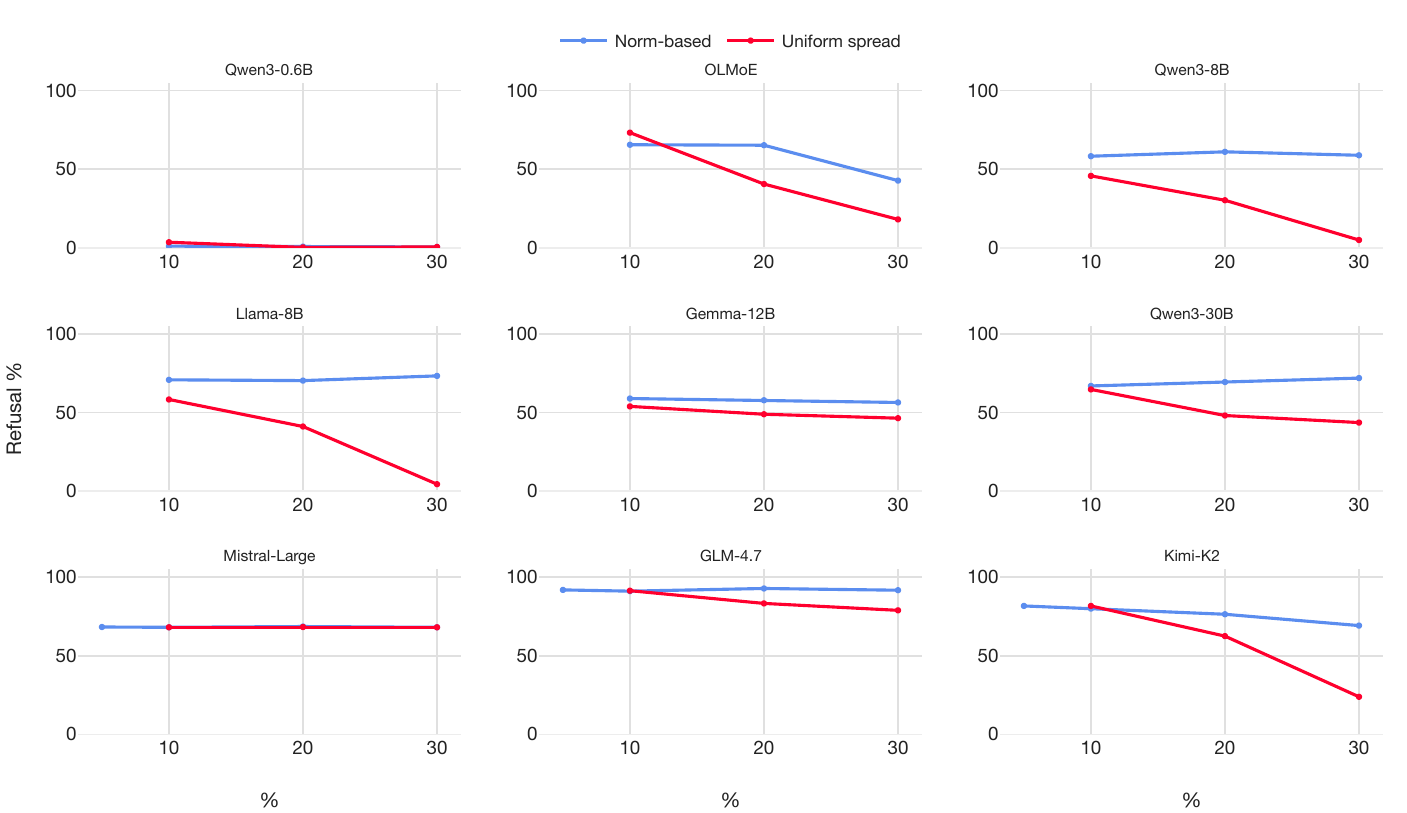}
\caption{Norm-based vs.\ uniform spread layer selection across 9 models. Uniform spread (red) consistently achieves equal or greater refusal reduction than norm-based (blue).}
\label{fig:app_layers}
\end{figure}

\textbf{Layer selection strategy} (Figure~\ref{fig:app_layers}) compares norm-based selection (targeting layers with highest refusal signal) to uniform spread (layers distributed across 25--95\% of depth). Uniform spread consistently matches or outperforms norm-based selection across all 9 tested models, with the gap reaching ${\sim}70$pp of additional refusal reduction at 30\% on the most susceptible models (e.g., Llama-3.1-8B, Qwen3-8B, Kimi-K2). On the resistant tier (Mistral-Large), neither method moves refusal at any percentile.

\section{Domain-Specificity Analysis}
\label{app:domain}

We extracted per-domain refusal directions for 5 representative models (Qwen3-8B, Mistral-Small-3.1-24B, Qwen3-30B-A3B, Llama-4-Scout, GLM-4.7) spanning dense and MoE architectures from 3.8B to 355B parameters. For each model, hidden states were extracted at all decoder layers, per-domain refusal directions computed as $\hat{\mathbf{r}}_{\ell}^{(d)} = \text{norm}(\text{mean}(\mathbf{h}_{\ell}^{\text{harmful},d}) - \text{mean}(\mathbf{h}_{\ell}^{\text{harmless}}))$, and pairwise cosine similarity averaged across layers in the 25--95\% depth range.

\begin{figure}[h]
\centering
\includegraphics[width=0.65\columnwidth]{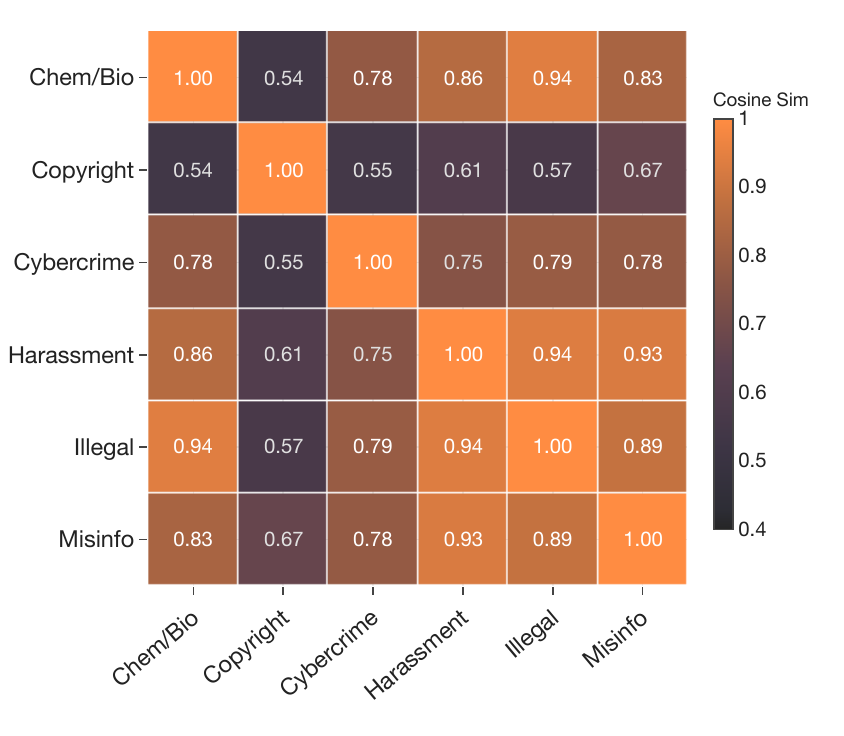}
\caption{Per-domain refusal direction similarity on the scientific benchmark, averaged across 5 models. Cybercrime and copyright show the most distinct refusal directions.}
\label{fig:sim_eval}
\end{figure}

\begin{figure}[h]
\centering
\includegraphics[width=0.65\columnwidth]{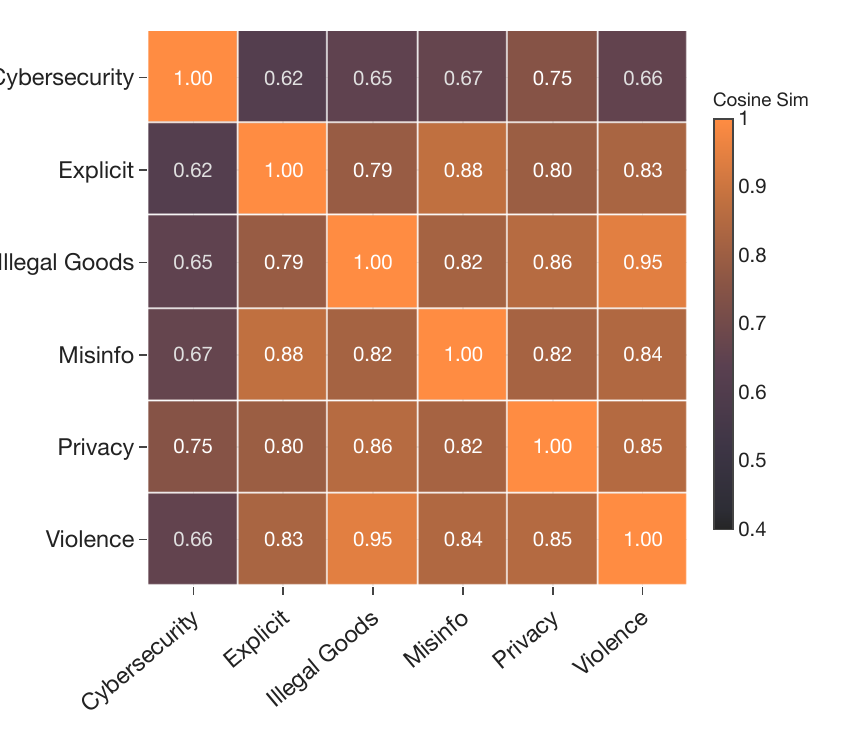}
\caption{Per-domain refusal direction similarity on the cross-evaluation dataset, averaged across 5 models. Cybersecurity has the most distinct refusal direction.}
\label{fig:sim_cross}
\end{figure}

Figure~\ref{fig:sim_cross} shows the aggregated similarity matrix on our cross-evaluation dataset (6 domains). Cybersecurity consistently has the lowest inter-domain similarity (0.56--0.70), confirming that its refusal direction is geometrically distinct. The maximum similarity pair is illegal goods $\leftrightarrow$ violence (0.94), suggesting these domains share nearly identical refusal geometry. Figure~\ref{fig:sim_eval} shows the same analysis on the scientific benchmark, where cybercrime and copyright emerge as the most distinct domains.

\section{Predictors of Susceptibility}
\label{app:predictors}

We analyzed several candidate predictors of abliteration susceptibility across all 24 models.

\begin{figure}[h]
\centering
\includegraphics[width=0.75\columnwidth]{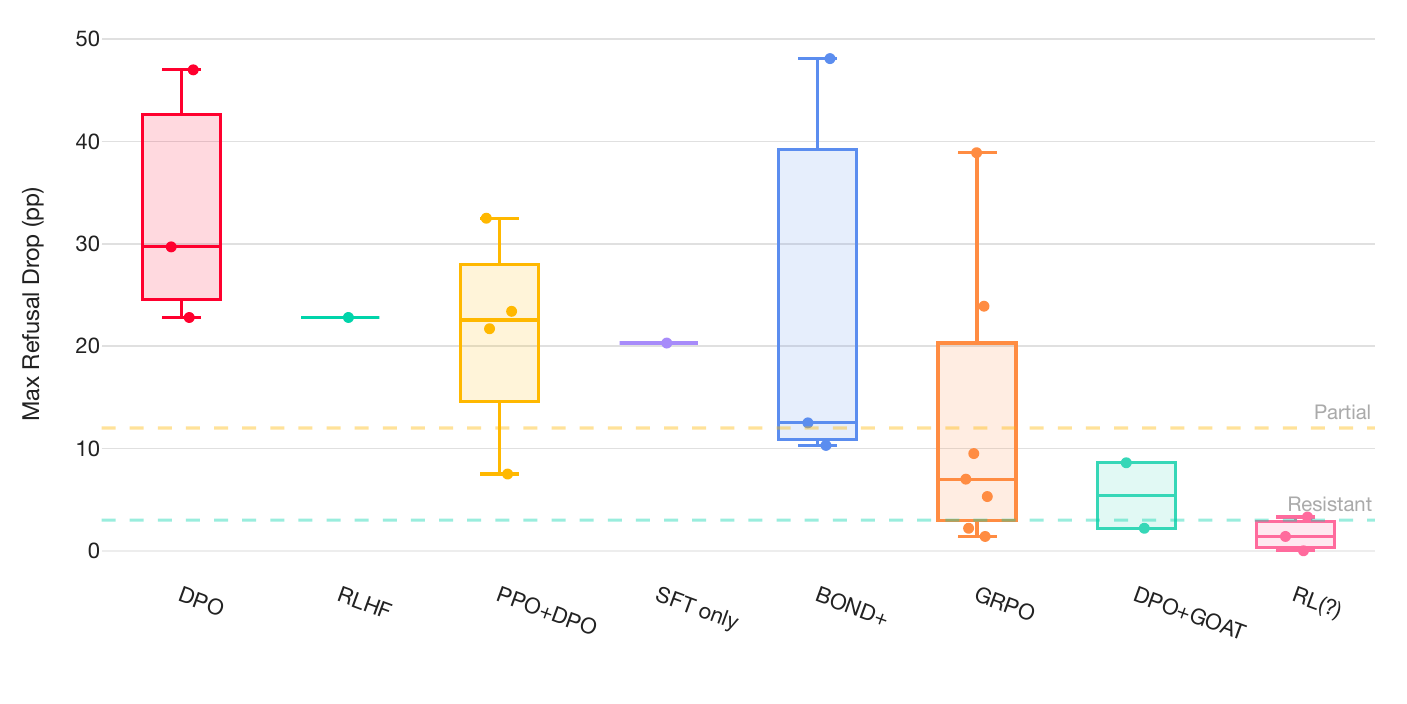}
\caption{Maximum refusal reduction by safety training method. DPO-based methods are most susceptible; GRPO and undisclosed RL methods confer near-resistance.}
\label{fig:app_safety}
\end{figure}

\textbf{Safety training method} (Figure~\ref{fig:app_safety}) is the strongest individual signal. DPO-based training produces the most susceptible models (median $\sim$30pp, all above the susceptibility threshold). RLHF-based and PPO+DPO methods cluster around 20--25pp. GRPO-based training (used by all Qwen3 variants) yields consistently low susceptibility (5--8pp), placing most models in the partial or resistant range. Models with undisclosed RL-based safety (Mistral-Small, Mistral-Large) are fully resistant ($<$1pp). Google's BOND+WARM+WARP method shows the widest variance, with Gemma-1B at $\sim$48pp and Gemma-27B at $\sim$13pp.

\begin{figure}[h]
\centering
\includegraphics[width=0.75\columnwidth]{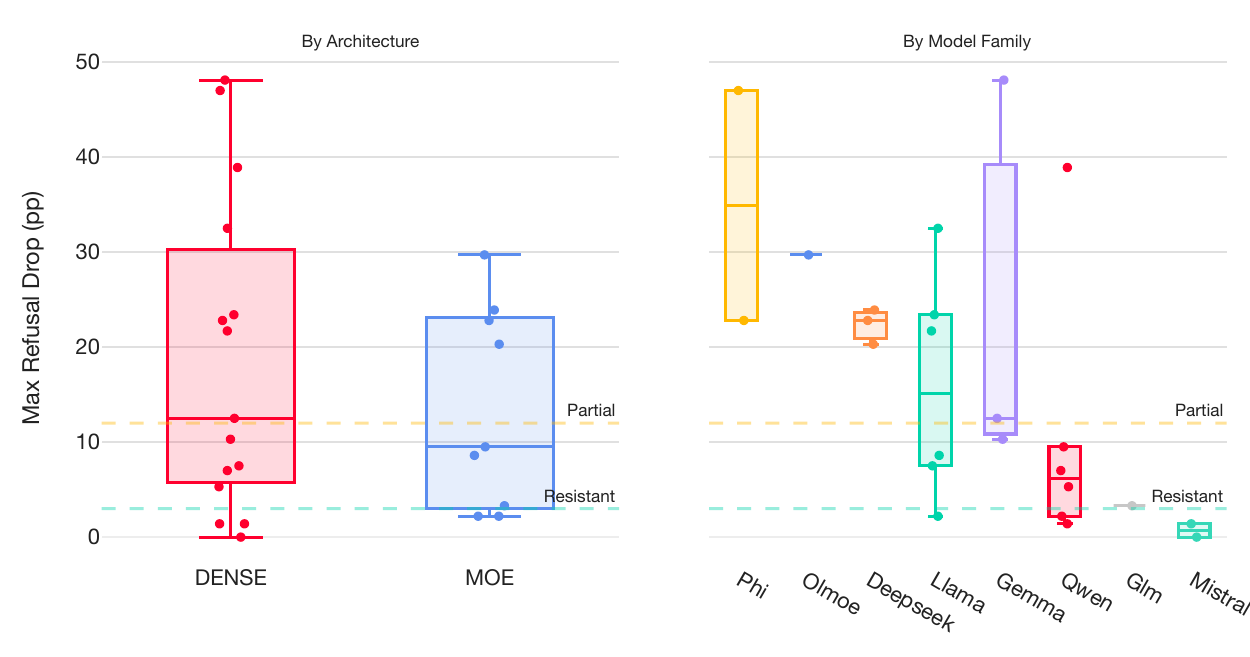}
\caption{Susceptibility by architecture type (left) and model family (right). Dense and MoE show comparable overall distributions, but family-level patterns emerge.}
\label{fig:app_arch}
\end{figure}

\textbf{Architecture and model family} (Figure~\ref{fig:app_arch}) show weaker but visible patterns. Dense transformers have a higher median susceptibility than MoE models, although both distributions overlap substantially. By family: Phi models are uniformly susceptible (all $>$20pp), Qwen spans the full range (1--39pp), Llama clusters in the mid-range, and Mistral is consistently resistant.

\end{document}